\documentclass[useAMS,usenatbib,dvipdfmx]{mn2e}
\usepackage[dvipdfmx]{graphicx}
\usepackage{array,booktabs,color}
\title[The Impact of Dust on Quasar Luminosity Functions]
  {The Impact of Dust in Host Galaxies on Quasar Luminosity Functions}
\author[H.~Shirakata et al.]
  {Hikari~Shirakata,$^1$\thanks{E-mail: shirakata@astro1.sci.hokudai.ac.jp} Takashi~Okamoto,$^1$ Motohiro~Enoki,$^2$ Masahiro~Nagashima,$^3$
  \newauthor
   Masakazu~A.~R.~Kobayashi,$^4$ Tomoaki~Ishiyama$^5$ and Ryu~Makiya$^6$\\
  $^1$Department of Cosmosciences, Hokkaido University, N10 W8, Kitaku, Sapporo, 060-0810, Japan\\
  $^2$Faculty of Business Administration, Tokyo Keizai University, Kokubunji, Tokyo, 185-8502, Japan\\
  $^3$Faculty of Education, Bunkyo University, Koshigaya, Saitama 343-8511, Japan\\
  $^4$Research Center for Space and Cosmic Evolution, Ehime University, Matsuyama, Ehime, 790-8577, Japan\\
  $^5$Center for Computational Sciences, University of Tsukuba, Tsukuba, Ibaraki, 305-8577, Japan\\
  $^6$Institute of Astronomy, The University of Tokyo, Mitaka, Tokyo, 181-0015, Japan}
\date{}
\pagerange{\pageref{firstpage}--\pageref{lastpage}} \pubyear{}

\begin{document}
\maketitle
\label{firstpage}

\begin{abstract}
We have investigated effects of dust attenuation on quasar luminosity 
functions at $z~\sim 2$ using a semi-analytic galaxy formation model combined with 
a large cosmological $N$-body simulation. 
We estimate the dust attenuation of quasars self-consistently 
with that of galaxies by considering the dust in their host bulges.
We find that the luminosity of the bright quasars is strongly dimmed by 
the dust attenuation, $\sim 2$~mag in the $B$-band.
Assuming the empirical bolometric corrections for active galactic nuclei (AGNs)
by Marconi et al., we find that this dust attenuation is too strong to
explain the $B$-band and X-ray quasar luminosity functions simultaneously.
We consider two possible mechanisms that weaken the dust attenuation.
As such a mechanism, we introduce a time delay for AGN activity, that is,
gas fueling to a central black hole starts some time after the beginning of the
starburst induced by a major merger. The other is the anisotropy
in the dust distribution. We find that in order to make the dust attenuation
of the quasars negligible, either the gas accretion into the black holes has to 
be delayed at least three times the dynamical timescale of their host bulges or the
dust covering factor is as small as $\sim$ 0.1.
\end{abstract}

\section{Introduction}

The mass of super massive black holes (SMBHs) correlates with 
the properties of their host galaxies,  such as stellar mass and velocity 
dispersion of their bulges \citep[e.g.,][]{Mg98, FM00,HR04,MM13}. 
These facts suggest that SMBHs have co-evolved with 
their host galaxies. 
The most straightforward way to understand the co-evolution is 
studying the properties of active galactic nuclei (AGNs), 
which emit light when materials get accreted by the SMBHs.  

Quasars are the brightest class of AGNs.
There have been many observational studies of AGN luminosity functions 
in several bands \citep[e.g.,][]{C09,C01,U14}.
Effects of dust attenuation have also been studied for 
various classes of AGNs, including quasars
\citep[e.g.,][]{SA88, GA07, LA07}.  
These studies suggest the existence of obscured quasars, which are 
hidden by the surrounding dust and can be observed only in X-ray. 
\citet{LA07} suggest that only 33\% of luminous AGNs
selected based on their mid-infrared colors
are unobscured (Type 1) quasars. 
This picture is in good accord with the simulation results by 
\citet{HO05a}; they suggest that quasars are buried in the dust 
in merger remnants and only become observable at their 
late evolutionary stage. 

Semi-analytic (SA) galaxy formation models are powerful tools to study rare objects like 
AGNs theoretically since they can calculate properties of individual 
galaxies in large cosmological $N$-body simulations.  
SA models have thus been widely used to investigate AGN properties 
\citep[e.g.,][]{KH00,E03,CA05A,MO07,LA08,MA08,FA12,HI12}. 
Many of them considered dust attenuation 
and attempted to reproduce observed luminosity functions
\citep[e.g.,][]{CA05A,FA12} by utilizing the empirical relations
\citep[e.g.,][]{U03,SZ04,BA05,HA08}.
None of them however consider effects of the dust attenuation on 
quasar luminosity self-consistently with properties of their host 
galaxies. 
In order to investigate effects of the dust on quasar luminosity 
functions self-consistently with galaxy formation, we employ an SA 
model, based on {\scriptsize $\nu$GC} 
\citep{nuGC} that is a galaxy formation model combined with 
high-resolution cosmological $N$-body simulations in a $\Lambda$-dominated 
cold dark matter ($\Lambda$CDM) universe. 
This model well reproduces the quasar luminosity functions 
around $z \sim 2$ and 
 naturally explains the cosmic downsizing of quasars 
\citep[][hereafter E14]{E14} together with many observed properties 
of galaxies \citep{nuGC}. 
Our model is identical to that used in E14, 
unless otherwise stated. 
In this {\it Letter}, we consider the dust in quasar host bulges, which 
attenuates the starlight from the bulge stars during starbursts; 
the same dust must attenuate the quasars, while this effect has 
been neglected in our previous studies \citep[][E14]{E03}. 

This {\it Letter} is organized as follows: in Section 2 we briefly 
review our models for the growth of the SMBHs and AGN luminosity. 
In Section 3 we present our results that indicate the quasars 
produced by our model are 
significantly attenuated by the dust in the host galaxies.
We also propose two mechanisms which make the dust attenuation
weak.
Finally we summarize and discuss our results in Section 4. 
 
\section{Modelling SMBH growth and quasar luminosity}

We create merging histories of dark matter haloes from a large 
cosmological $N$-body simulation. 
The baryonic processes in each halo are calculated by 
using simple, parametric forms of equations. 
Throughout this {\it Letter}, we assume a $\Lambda$CDM universe with 
the following parameters: 
$\Omega_{0}=0.2725$,  $\Omega_\Lambda=0.7275$, 
$\Omega_\mathrm{b} = 0.0455$, 
$\sigma_8 = 0.807$, $n_\mathrm{s} = 0.961$, and a Hubble constant of 
$H_0 = 100~h~\mathrm{km}~\mathrm{s}^{-1}$~Mpc$^{-1}$, where $h=0.702$ 
\citep{wmap7}.
Our cosmological $N$-body simulation contains $2048^{3}$ particles within
a co-moving box size of 280$~h^{-1}$~Mpc. 
The minimum halo mass is $7.72 \times 10^{9}~h^{-1}M_\odot$.
Further details can be found in E14 and \citet{I14}.
In this section, we briefly describe the 
processes that are most relevant to our study.  
 
\subsection{Growth of SMBHs}

Galaxies in a common dark matter halo sometimes merge together by 
dynamical friction or random collisions. 
When two or more galaxies merge, their central SMBHs also 
coalesce into a single SMBH. 
We define a major merger as
an event where the mass ratio of the secondary
progenitor to the primary one is larger than 0.4. 
In this case, the cold gas in the merger remnant is converted into 
stars with a short timescale and a fraction of it gets accreted by
the SMBH. 
The mass of the cold gas accreted by the SMBH, $M_{\mathrm{acc}}$, is 
given by 
\begin{equation}
M_{\mathrm{acc}}=f_{\mathrm{BH}}\Delta M_{*,\mathrm{burst}},
\end{equation}
where $\Delta M_{*,\mathrm{burst}}$ is the mass of stars that form 
during the starburst and $f_{\mathrm{BH}}$ is a free parameter whose  
value is set to 0.01 in order to reproduce the observed 
relationship between bulge mass and black hole mass at $z = 0$ \citep{HR04,MM13}. 

\subsection{AGN luminosity} \label{sec:2}

Assuming that a fixed fraction of the rest mass energy of the 
accreted cold gas by an SMBH is radiated  
and that an AGN light curve has an exponentially declining form 
as in E14, 
we obtain the bolometric luminosity of an AGN at 
time $t$ after a major merger as
\begin{equation}
  L_\mathrm{bol}(t) = \frac{\epsilon_\mathrm{bol} M_\mathrm{acc} c^2}{t_\mathrm{life}} \exp(-t/t_\mathrm{life}), 
  \label{eq:lightcurve}
\end{equation}
where $\epsilon_\mathrm{bol}$ is the bolometric radiative efficiency,
$t_\mathrm{life}$ is the AGN lifetime, and $c$ is the speed of light. 
We suppose that the AGN lifetime is proportional to the dynamical 
timescale of the bulge\footnote{E14 assumed that the AGN lifetime  
is proportional to the dynamical timescale of the host {\it halo}.
Although this modification does not affect the results in E14 
qualitatively, the values of $\epsilon_B$ and $f_\mathrm{life}$ 
must be adjusted to obtain equivalent results to E14. 
}, 
$t_\mathrm{dyn}$, 
as $t_\mathrm{life} = f_\mathrm{life} t_\mathrm{dyn}$.

\citet{MA04} propose the bolometric corrections for hard X-ray (2--10 keV), 
soft X-ray (0.5--2 keV) and $B$-band 
(hereafter ``Marconi relations''). 
We obtain hard X-ray luminosity, $L_{X}$ and $B$-band luminosity, $L_{B}$
from Marconi relations as
\begin{eqnarray}
  \lefteqn{\log [L/L \textrm{(2--10 keV)}] = 1.54 + 0.24\mathcal{L} + 0.012\mathcal{L}^{2} - 0.0015\mathcal{L}^{3},}  \nonumber \\
  \lefteqn{\log (L/\nu_{B}L_{\nu_{B}}) = 0.80 - 0.067\mathcal{L} + 0.017\mathcal{L}^{2} - 0.0023\mathcal{L}^{3},}
\end{eqnarray}
where $\mathcal{L} = (\log L - 12)$ and $L$ is the intrinsic 
bolometric luminosity in units of $L_{\odot}$. 

We chose the values of $\epsilon_\mathrm{bol}$ and $f_\mathrm{life}$ to match 
the observed AGN luminosity functions in hard X-ray
at $z \sim 2$, 
at which the comoving number density of the quasars is the highest.
We employ $\epsilon_\mathrm{bol} = 0.1$ and $f_\mathrm{life} = 1.5$
throughout this {\it Letter}.
If $t_\mathrm{life}$ is shortened, then the faint end slope
of the AGN luminosity function is shallower and the number
of bright AGNs increases 
as mentioned in \cite{cat01} and \cite{E03}.
We assume that AGNs with $N_\mathrm{H} > 10^{24}~\textrm{cm}^{-2}$
are Compton-thick and that they will be missed
even by hard X-ray surveys.
\subsection{Dust attenuation}

The dust attenuation of galaxies has been an important ingredient in 
SA models \citep[e.g.,][]{CO00, nuGC}. 
In our model, the optical depth of internal dust in a merger remnant 
at the $V$-band is assumed to be 
\begin{equation}
  \tau_V = \tau_{V0} \frac{(M_\mathrm{cold}/\mathrm{M}_\odot)
  (Z_\mathrm{cold}/Z_\odot)}{(r_{e}/\mathrm{kpc})^{2}}(1+z)^{-\gamma},
\label{eq:optdepth}
\end{equation}
where $\tau_{V0} = 2.5 \times 10^{-9}$ is the 
proportionality constant, $M_{\mathrm{cold}}$ is the total cold gas mass in the merger 
remnant, $Z_{\mathrm{cold}}$ is the metallicity of the cold gas, 
$r_{e}$ is the effective radius of the merger remnant (bulge), 
and $\gamma$ is a free parameter chosen to predict a consistent number 
of high redshift galaxies with observations. 
As in \citet{nuGC}, we set $\gamma  = 1$.
Although \cite{CO00} assumed local extinction \citep{SM79}, $N_{H}/E(B-V) = 5.8 \times 10^{21}~\textrm{cm}^{-2}$,
we employ half of the value
because we adopt the value of the chemical yield 
twice as large as that in \cite{CO00} in order to reproduce observed color  
of elliptical galaxies \citep{nuGC}.
We employ the extinction curve by \citet{Cal00}, $R_{V} = 4.01$,
and obtain $\tau_{B0} = 3.1 \times 10^{-9}$.
The time evolution of $M_\mathrm{cold}$ and $Z_\mathrm{cold}$ 
after a major merger is also given in \citet{nuGC}.

An AGN must also be subject to the attenuation by the same dust 
that attenuates the merger remnant, 
since the AGN is embedded in the very centre of it. 
While we adopt a slab model for galaxies, we employ a screen model 
for AGNs. We also halve the optical depth obtained by 
Equation~(\ref{eq:optdepth}) to compute the attenuation of 
AGNs because an AGN is located at the centre of its host bulge. 

The other important physical parameter is the star formation 
timescale for a starburst, $\tau_\mathrm{burst}$, since 
the cold gas is exhausted by this timescale. 
We assume that this timescale scales with the dynamical timescale 
of a bulge as $\tau_\mathrm{burst} = 0.5 t_\mathrm{dyn}$. 
The proportionality constant has been chosen to explain the colour 
of high redshift starburst galaxies \citep{nuGC}. 

\subsection{Models}
\label{sec:models}
In this {\it Letter}, we consider three models regarding 
the dust attenuation of AGNs and investigate quasar luminosity functions.
In the first model, 
and the first model only, we ignore the dust attenuation.
In this case the luminosity of an AGN is fully specified by
Equation~(\ref{eq:lightcurve}).
This model is an analogue of the one presented in E14 and 
we call this model ``E14 model''.
We note that this E14 model
is different from the original E14 model in the treatment of dust
attenuation; in the original E14 model, dust attenuation is effectively
included since $\epsilon_B$ is treated as a free parameter.

In E14 model, we assume that the accretion onto an SMBH and 
a starburst begin immediately after the occurrence of a major merger.
It may however take some time before the gas reaches to the galactic
centre after a major merger. 
We hence consider the second model in which gas accretion onto an SMBH 
starts $\tau_\mathrm{delay} = \nu_\mathrm{delay} t_\mathrm{dyn}$ after 
a major merger; 
doing this must reduce the dust attenuation of AGNs.
This delay is firstly introduced in this {\it Letter}. 
We refer to this model as ``delay model'' and
present the results of $\nu_{\mathrm{delay}} = 1$ and 3.

In the third model, we vary the dust covering factor (CF).
We present the results of CF = 0.1, 0.5, 0.75, 0.9,
while CF = 1 is assumed in E14  and delay models. 
We dub this model ``CF model''.

Quasars are required to be brighter than their host galaxies
in our definition.

\section{Results}
In order to focus on these effects,
hereafter we show the luminosity functions of the quasars only 
at $z~\sim 2$.
Note that our model overpredicts the number of bright quasars
in the $B$-band about an order of magnitude at $z < 0.5$. 
We leave this issue for future studies.

\begin{figure}
  \includegraphics[width=\linewidth]{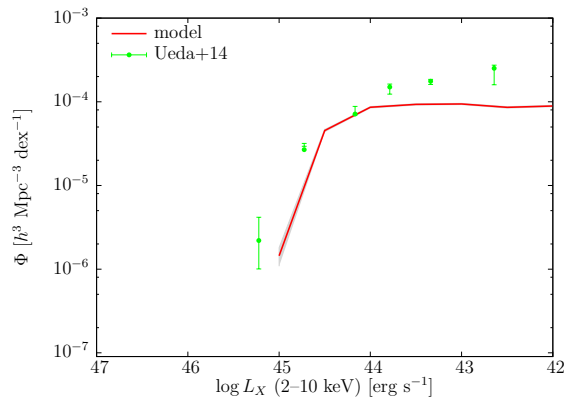}
  \caption{AGN luminosity functions in hard X-ray.
    We set $\epsilon_\mathrm{bol}$ and $f_\mathrm{life}$ 
    so that the model reproduces the observed 
    luminosity function obtained by \citet{U14} (green points).
    Note that the model include only AGNs that are 
    triggered by major mergers.
    Hence it is not surprising that it misses the less powerful 
    Seyfert population.
    Red line shows the model result. 
    The shaded region indicates 1$\sigma$ error.
    }
  \label{fig:ALF-X}
\end{figure}
In Fig.~\ref{fig:ALF-X}, 
we present AGN luminosity functions 
in hard X-ray at $z~\sim 2$. (red solid line).
We presume that the dust attenuation is negligible
in hard X-ray.
Our result is broadly consistent with the observed luminosity
function. Since, in our model, we only consider quasar-like AGNs,
i.e. AGNs induced by major mergers, it is natural that our model underpredicts
the number of faint AGNs.
The brightest end is highly uncertain due to the cosmic variance.

\begin{figure}
  \includegraphics[width=\linewidth]{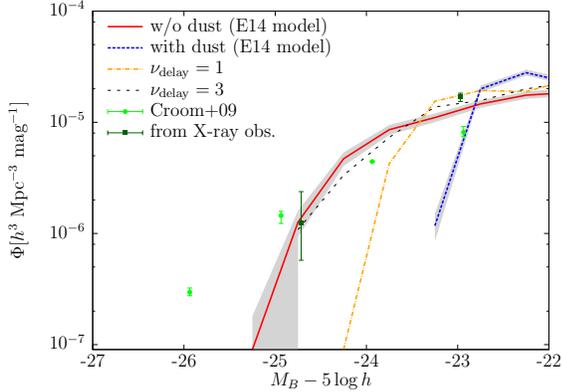}
  \caption{Quasar luminosity functions in $B$-band
  at $z~\sim 2$. The luminosity function of E14 model is indicated
  by the red solid line in which we ignore the dust attenuation.
  We also show the result by the same parameter set
  as E14 model but with the dust attenuation (blue doted line).
  The shaded regions indicate 1$\sigma$ error.
  The observational estimation
  is indicated by the green filled circles \citep{C09}.
  The dark-green filled squares indicate the luminosity function that is 
  converted from the hard X-ray luminosity function obtained by \citet{U14}
  using Marconi relations. These points are corresponding to 
  the first and second brightest points in Fig.~\ref{fig:ALF-X}.
  We also show the results of delay model 
  with $\nu_\mathrm{delay} = 1$ (orange dot dashed line) and
  with $\nu_\mathrm{delay} = 3$ (black dashed line).}
  \label{fig:delayedLF}
\end{figure}

In Fig.~\ref{fig:delayedLF}, we present quasar luminosity functions 
in $B$-band at $z~\sim 2$ predicted by E14 model (red solid line).
We also show the luminosity function that is converted 
from the observed AGN luminosity function in hard X-ray
obtained by \citet{U14} using Marconi relations (dark-green filled squares)
together with an observed $B$-band luminosity function (\citealt{C09}, green filled circles).
These two luminosity functions are consistent.
This suggests that if Marconi relations are appropriate, 
the dust attenuation of quasars 
in $B$-band must be negligible because Marconi relation gives 
intrinsic $B$-band luminosity. 
We also plot the luminosity function by the same parameter set as E14 model 
but with the dust attenuation (blue doted line).
The dust distribution that we have assumed here is 
consistent with the dust model assumed for the host bulges.
We find that the luminosity of the brightest quasar becomes fainter by about 2~mag
and that is inconsistent with the observed quasar luminosity function.
If we ignore the dust attenuation, we can roughly reproduce observed quasar luminosity
function.
Hence in order to reproduce the X-ray and $B$-band luminosity functions simultaneously, 
we should introduce some mechanisms that weaken the dust attenuation.

We now investigate effects of the delayed accretion onto SMBHs (delay model). 
In Fig.~\ref{fig:delayedLF}, we compare luminosity functions with 
$\nu_\mathrm{delay} = 0$ (E14 model), 1, and 3. 
We find that this effect is larger for the bright end. 
Thanks to the smaller amount of the cold gas due to star formation 
and feedback,
the dust attenuation becomes weak.
Since the red solid line (E14 model) and black dashed line 
($\nu_\mathrm{delay} = 3$)
in Fig.~\ref{fig:delayedLF} are indistinguishable, 
we conclude that, in order to make the dust attenuation to be 
negligible, the gas accretion onto an SMBH has to wait  
$\sim 3 t_\mathrm{dyn}$ after a major merger.

\begin{figure}
  \includegraphics[width=\linewidth]{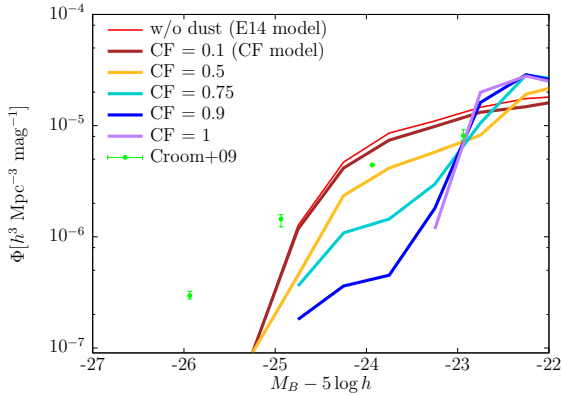}
  \caption{Quasar luminosity functions of CF model. 
    Red solid line shows the quasar luminosity function without dust attenuation.
    Other 5 thick lines show that of with CF = 0.1 (brown solid line), 0.5 (orange solid line),
    0.75 (cyan solid line), 0.9 (blue solid line) and 1 (purple solid line);
    the smaller CF corresponds to the brighter bright end.}
  \label{fig:CF}
\end{figure}
Finally, we have tested the CF model
in Fig.~\ref{fig:CF}. We find that the shape of the quasar luminosity 
functions vary according to the CF, and that the dust attenuation can be
negligible if CF $\sim 0.1$. 
The dust distribution therefore plays an important role 
for dust attenuation as mentioned in \cite{CA05B}.

\section{Summary and discussion} 

We have analysed the quasars obtained by an SA galaxy 
formation model of E14, which naturally 
explains the anti-hierarchical evolution of the AGNs, 
to study the dust attenuation of the quasars self-consistently 
with galaxy formation. 

We find that bright quasars suffer significant dust 
attenuation, $\sim 2$~mag in $B$-band, 
if the accretion onto a central black hole occurs simultaneously with a 
starburst. 
Hence we should introduce some mechanisms which weaken the dust attenuation
in order for our model to reproducde observed quasar luminosity functions,
both in the $B$-band and X-ray.

The dust attenuation is weakened if the accretion onto 
an SMBH starts some time after a major merger since  
the amount of the cold gas and hence the dust in the host 
bulge is reduced by star formation and feedback.   
We take this process into account for the first time in SA galaxy formation models.
We find that the delay must be at least 
$\sim 3 t_\mathrm{dyn}$ in order for the dust attenuation to 
be completely negligible. 
This result suggests that the SMBH growth might significantly lag
behind the peak star formation activity in the host.

The dust distribution that we have assumed in E14 and delay models
is consistent with the dust model assumed for the host bulges.
It is however possible that the dust attenuates the light only along 
certain lines of sight.
We find that if CF $\sim$ 0.1, the dust attenuation becomes negligible.
The dust covering factor may play an important role for dust attenuation
of AGNs in late type galaxies \citep{R06}.
We however consider gas rich major mergers as a trigger of AGN activities, 
and therefore CF $\sim$ 1 might be plausible \citep{CA05B}.

AGN feedback, in particular so-called quasar mode feedback 
\citep[e.g.,][]{DI05, HO05a},  
may also affect the dust attenuation by blowing  
away the cold gas from a bulge \citep{HO05a}. 
Including this effect however changes galaxy properties as well, 
and thus we have to weaken the stellar feedback to obtain the 
equivalent results for galaxies. 
We hence speculate that inclusion of this feedback will not change our 
results much.

We conclude that the attenuation by the dust in the host galaxies
is so large that we cannot ignore it for modelling quasars.
We have not considered nuclear dust in this {\it Letter}, which 
might be more important than the dust in the host galaxies 
because of its higher column density. The role of the AGN feedback
is likely to be more important for the nuclear dust than the galactic
one \citep{HO05a}.
To consider the nuclear dust, we however have to model the nuclei more in 
detail and we leave it for future studies.

\section*{Acknowledgements}
We appreciate the detailed review 
and useful suggestions of anonymous
referee that have improved our paper.
We would like to thank T. Nagao, M. Akiyama, T. Kawaguchi,
Y. Matsuoka, M. Brown and Y. Toba 
for helpful comments on observational data.
We also thank K. Wada, T. Kozasa, K. Ohsuga, T. Oogi and K. Hiura
for useful discussion.
T.O. has been financially supported by 
Japan Society for the Promotion 
of Science (JSPS) Grant-in-Aid for Young Scientists (B: 24740112).
M.N. has been supported by the Grant-in-Aid (No. 25287049) from the Ministry
of Education, Culture, Sports, Science, and Technology (MEXT) of
Japan.
T.I. has been supported by MEXT HPCI STRATEGIC PROGRAM and MEXT/JSPS KAKENHI
(No. 24740115).
R.M. has been supported by the Grant-in-Aid for JSPS Fellows.

\label{lastpage}

\begin{thebibliography}{}
\bibitem[\protect\citeauthoryear{Barger et al.}{2005}]{BA05} Barger A.~J., Cowie L.~L., Mushotzky R.~F., Yang Y., Wang W.-H., Steffen A.~T., Capak P., 2005, AJ, 129, 578 
\bibitem[\protect\citeauthoryear{Calzetti et al.}{2000}]{Cal00} Calzetti D., Armus L., Bohlin R.~C., Kinney A.~L., Koornneef J., Storchi-Bergmann T., 2000, ApJ, 533, 682
\bibitem[\protect\citeauthoryear{Cattaneo}{2001}]{cat01} Cattaneo A., 2001, MNRAS, 324, 128 
\bibitem[\protect\citeauthoryear{Cattaneo et al.}{2005a}]{CA05A} Cattaneo A., Blaizot J., Devriendt J., Guiderdoni B., 2005a, MNRAS, 364, 407
\bibitem[\protect\citeauthoryear{Cattaneo et al.}{2005b}]{CA05B} Cattaneo A., Combes F., Colombi S., Bertin E., Melchior A.-L., 2005b, MNRAS, 359, 1237
%\bibitem[\protect\citeauthoryear{Chen et al.}{2015}]{C15} Chen C.-T.~J., et al., 2015, arXiv, arXiv:1501.04959
\bibitem[\protect\citeauthoryear{Cole et al.}{2000}]{CO00} Cole S., Lacey C.~G., Baugh C.~M., Frenk C.~S., 2000, MNRAS, 319, 168
\bibitem[\protect\citeauthoryear{Croom et al.}{2001}]{C01} Croom S.~M., Smith R.~J., Boyle B.~J., Shanks T., Loaring N.~S., Miller L., Lewis I.~J., 2001, MNRAS, 322, L29
\bibitem[\protect\citeauthoryear{Croom et al.}{2009}]{C09} Croom S.~M., et al., 2009, MNRAS, 399, 1755 

\bibitem[\protect\citeauthoryear{Di Matteo, Springel, \& Hernquist}{2005}]{DI05} Di Matteo T., Springel V., Hernquist L., 2005, Natur, 433, 604 
%\bibitem[\protect\citeauthoryear{Elvis et al.}{1994}]{E94} Elvis M., et al., 1994, ApJS, 95, 1
\bibitem[\protect\citeauthoryear{Enoki et al.}{2014}]{E14} Enoki M., Ishiyama T., Kobayashi M.~A.~R., Nagashima M., 2014, ApJ, 794, 69 
\bibitem[\protect\citeauthoryear{Enoki, Nagashima, \& Gouda}{2003}]{E03} Enoki M., Nagashima M., Gouda N., 2003, PASJ, 55, 133
\bibitem[\protect\citeauthoryear{Fanidakis et al.}{2012}]{FA12} Fanidakis N., et al., 2012, MNRAS, 419, 2797
\bibitem[\protect\citeauthoryear{Ferrarese \& Merritt}{2000}]{FM00} Ferrarese L., Merritt D., 2000, ApJ, 539, L9
%\bibitem[\protect\citeauthoryear{Francis, Hooper, \& Impey}{1993}]{F93} Francis P.~J., Hooper E.~J., Impey C.~D., 1993, AJ, 106, 417
\bibitem[\protect\citeauthoryear{Gaskell \& Benker}{2007}]{GA07} Gaskell C.~M., Benker A.~J., 2007, arXiv, arXiv:0711.1013
\bibitem[\protect\citeauthoryear{H{\"a}ring \& Rix}{2004}]{HR04} H{\"a}ring N., Rix H.-W., 2004, ApJ, 604, L89
\bibitem[\protect\citeauthoryear{Hasinger}{2008}]{HA08} Hasinger G., 2008, A\&A, 490, 905
\bibitem[\protect\citeauthoryear{Hirschmann et al.}{2012}]{HI12} Hirschmann M., Somerville R.~S., Naab T., Burkert A., 2012, MNRAS, 426, 237
%\bibitem[\protect\citeauthoryear{Hopkins, Richards, \& Hernquist}{2007}]{HO07} Hopkins P.~F., Richards G.~T., Hernquist L., 2007, ApJ, 654, 731
\bibitem[\protect\citeauthoryear{Hopkins et al.}{2005}]{HO05a} Hopkins P.~F., Hernquist L., Cox T.~J., Di Matteo T., Martini P., Robertson B., Springel V., 2005, ApJ, 630, 705
%\bibitem[\protect\citeauthoryear{Hopkins et al.}{2005b}]{Ho05b} Hopkins P.~F., Hernquist L., Cox T.~J., Di Matteo T., Robertson B., Springel V., 2005b, ApJ, 630, 716
%\bibitem[\protect\citeauthoryear{Hopkins et al.}{2005c}]{HO05c} Hopkins P.~F., Hernquist L., Martini P., Cox T.~J., Robertson B., Di Matteo T., Springel V., 2005c, ApJ, 625, L71
%\bibitem[\protect\citeauthoryear{Hopkins et al.}{2004}]{HO04} Hopkins P.~F., et al., 2004, AJ, 128, 1112
%\bibitem[\protect\citeauthoryear{Ikeda et al.}{2012}]{I12}Ikeda H., et al., 2012, ApJ, 756, 160 
%\bibitem[\protect\citeauthoryear{Ikeda et al.}{2011}]{I11} Ikeda H., et al., 2011, ApJ, 728, L25 
\bibitem[\protect\citeauthoryear{Ishiyama et 
  al.}{2014}]{I14} Ishiyama T., Enoki M., Kobayashi M.~A.~R., Makiya R., Nagashima M., Oogi T., 2014, arXiv, arXiv:1412.2860 
\bibitem[\protect\citeauthoryear{Kauffmann \& Haehnelt}{2000}]{KH00} Kauffmann G., Haehnelt M., 2000, MNRAS, 311, 576
\bibitem[\protect\citeauthoryear{Komatsu et al.}{2011}]{wmap7} Komatsu E., et al., 2011, ApJS, 192, 18
%\bibitem[\protect\citeauthoryear{Korista et al.}{1997}]{K97} Korista K., Baldwin J., Ferland G., Verner D., 1997, ApJS, 108, 401
\bibitem[\protect\citeauthoryear{Lacy et al.}{2007}]{LA07} Lacy M., Petric A.~O., Sajina A., Canalizo G., Storrie-Lombardi L.~J., Armus L., Fadda D., Marleau F.~R., 2007, AJ, 133, 186
%\bibitem[\protect\citeauthoryear{Lagos et al.}{2011}]{LA11} Lagos C.~D.~P., Padilla N.~D., Strauss M.~A., Cora S.~A., Hao L., 2011, MNRAS, 414, 2148 
\bibitem[\protect\citeauthoryear{Lagos, Cora, \& Padilla}{2008}]{LA08} Lagos C.~D.~P., Cora S.~A, Padilla N.~D., 2008, MNRAS, 388, 587
\bibitem[\protect\citeauthoryear{Magorrian et al.}{1998}]{Mg98} Magorrian J., et al., 1998, AJ, 115, 2285
\bibitem[\protect\citeauthoryear{Marconi et al.}{2004}]{MA04} Marconi A., Risaliti G., Gilli R., Hunt L.~K., Maiolino R., Salvati M., 2004, MNRAS, 351, 169
\bibitem[\protect\citeauthoryear{Marulli et al.}{2008}]{MA08} Marulli F., Bonoli S., Branchini E., Moscardini L., Springel V., 2008, MNRAS, 385, 1846
\bibitem[\protect\citeauthoryear{McConnell \& Ma}{2013}]{MM13} McConnell N.~J., Ma C.-P., 2013, ApJ, 764, 184  
\bibitem[\protect\citeauthoryear{Monaco, Fontanot, \& Taffoni}{2007}]{MO07} Monaco P., Fontanot F., Taffoni G., 2007, MNRAS, 375, 1189
%\bibitem[\protect\citeauthoryear{Nagashima \& Gouda}{2003}]{2003naoj.book...16N} Nagashima M., Gouda N., 2003, naoj.book, 16 
\bibitem[\protect\citeauthoryear{Nagashima et al.}{2005}]{nuGC} Nagashima M., Yahagi H., Enoki M., Yoshii Y., Gouda N., 2005, ApJ, 634, 26 
%\bibitem[\protect\citeauthoryear{Nagashima \& Yoshii}{2004}]{NY04} Nagashima M., Yoshii Y., 2004, ApJ, 610, 23
\bibitem[\protect\citeauthoryear{Rigby et al.}{2006}]{R06} Rigby J.~R., Rieke G.~H., Donley J.~L., Alonso-Herrero A., P{\'e}rez-Gonz{\'a}lez P.~G., 2006, ApJ, 645, 115 
\bibitem[\protect\citeauthoryear{Sanders et al.}{1988}]{SA88} Sanders D.~B., Soifer B.~T., Elias J.~H., Neugebauer G., Matthews K., 1988, ApJ, 328, L35
\bibitem[\protect\citeauthoryear{Savage \& Mathis}{1979}]{SM79} Savage B.~D., Mathis J.~S., 1979, ARA\&A, 17, 73
\bibitem[\protect\citeauthoryear{Szokoly et al.}{2004}]{SZ04} Szokoly G.~P., et al., 2004, ApJS, 155, 271 
\bibitem[\protect\citeauthoryear{Ueda et al.}{2014}]{U14} Ueda Y., Akiyama M., Hasinger G., Miyaji T., Watson M.~G., 2014, ApJ, 786, 104
\bibitem[\protect\citeauthoryear{Ueda et al.}{2003}]{U03} Ueda Y., Akiyama M., Ohta K., Miyaji T., 2003, ApJ, 598, 886 
%\bibitem[\protect\citeauthoryear{Zamorani et al.}{1981}]{Z81} Zamorani G., et al., 1981, ApJ, 245, 357
\end{thebibliography}
\end{document}